\documentclass{nature}

\usepackage{epsfig}

\def\arcmin{^{\prime}}

\def\gtrsim{\mathrel{\hbox{\rlap{\hbox{\lower4pt\hbox{$\sim$}}}\hbox{$>$}}}}
\def\lesssim{\mathrel{\hbox{\rlap{\hbox{\lower4pt\hbox{$\sim$}}}\hbox{$<$}}}}

\def\vkm{km s$^{-1}$}

\def\degree{$^\circ$}
\def\arcs#1{$#1''$}
\def\arcsa#1#2{$#1^{\prime\prime}_{^\textrm{.}}#2$}

\def\solarmass{$M_\odot$}

\def\solarlum{$L_\odot$}

\def\mJyb{mJy beam$^{-1}$}

\def\cmc{cm$^{-3}$}

\def\degr{\hbox{$^\circ$}}

\def\ra#1#2#3#4{#1^\mathrm{h} #2^\mathrm{m} #3^\mathrm{s}_{^\textrm{.}} #4}
\def\dec#1#2#3#4{#1\degr #2\arcmin #3^{\prime\prime}_{^\textrm{.}}#4}

\def\mH2{m_{\textrm{\scriptsize H}_2}}
\def\Ro{R_\textrm{\scriptsize 0}}
\def\ro{r_\textrm{\scriptsize 0}}

\def\To{T_\textrm{\scriptsize 0}}
\def\no{n_\textrm{\scriptsize 0}}

\def\H2{H$_2$}
\def\N2HP{N$_2$H$^+$}

\def\NH3{NH$_3$}

\def\SOt{$N_J=8_9-7_8$}

\def\putfig#1#2#3{\epsfig{scale=#1,angle=#2,figure=#3}}
\def\leftblank#1{}

\newcounter{mfigure}[section]
\newenvironment{mfigure}[1][]{\refstepcounter{mfigure}\par\medskip
   \noindent \textbf{Figure~\themfigure. #1} \rmfamily}{\medskip}

\title{A Rotating Protostellar Jet Launched from the Innermost Disk of HH 212}

\author{Chin-Fei Lee$^{1,2*}$, Paul. T.P Ho$^{1,3}$, Zhi-Yun Li$^{4}$ , Naomi
Hirano$^{1}$, Qizhou Zhang$^{3}$, \& Hsien Shang$^{1}$}

\begin{document}

\maketitle

\begin{affiliations}
 \item Academia Sinica Institute of Astronomy and Astrophysics, P.O. Box 23-141, Taipei 106, Taiwan
 \item Graduate Institute of Astronomy and Astrophysics, National Taiwan 
   University, No.  1, Sec.  4, Roosevelt Road, Taipei 10617, Taiwan
 \item Harvard-Smithsonian Center for Astrophysics, 60 Garden Street, Cambridge, MA 02138
 \item Astronomy Department, University of Virginia, Charlottesville, VA 22904
\end{affiliations}


\begin{abstract}

The central problem in forming a star is the angular momentum in the
circumstellar disk which prevents material from falling into the central
stellar core.  An attractive solution to the ``angular momentum problem"
appears to be the ubiquitous (low-velocity and poorly-collimated) molecular
outflows and (high-velocity and highly-collimated) protostellar jets
accompanying the earliest phase of star formation that remove angular
momentum at a range of disk radii\cite{Frank2014}.  Previous observations
suggested that outflowing material carries away the excess angular momentum
via magneto-centrifugally driven winds from the surfaces of circumstellar
disks down to $\sim$ 10 AU
scales\cite{Greenhill1998,Launhardt2009,Matthews2010,Greenhill2013,Bjerkeli2016},
allowing the material in the outer disk to transport to the inner disk. 
Here we show that highly collimated protostellar jets remove the residual
angular momenta at the $\sim$ 0.05 AU scale, enabling the material in the
innermost region of the disk to accrete toward the central protostar. 
This is supported by the rotation of the jet measured down to $\sim$ 10 AU
from the protostar in the HH 212 protostellar system. The measurement
implies a jet launching radius of $\sim 0.05_{-0.02}^{+0.05}$~AU on the
disk, based on the magneto-centrifugal theory of jet production, which
connects the properties of the jet measured at large distances to those at
its base through energy and angular momentum
conservation\cite{Anderson2003}.

\end{abstract}
 
\clearpage


Molecular outflows and protostellar jets together may solve the angular
momentum problem in disk accretion.  Molecular outflows have been found to
trace low-velocity ($\lesssim 20$ \vkm{}) extended winds coming out of the
disks down to $\sim$ 10 AU
scales\cite{Greenhill1998,Launhardt2009,Matthews2010,Greenhill2013,Bjerkeli2016}. 
They are found to be rotating and thus can remove angular momentum from the
disks in the outer part, allowing the disk material in the wind-launching
region to accrete.  Protostellar jets are much more collimated and denser
with a much higher velocity ($\gtrsim 100$ \vkm{}) than the molecular
outflows.  They are ejected out along the rotational axis from the innermost
part of the disks at less than 1 AU scale, and thus are expected to carry
away angular momentum from there, allowing the disk material to fall onto
the central protostars from the disk truncation radii\cite{Frank2014}. 
Tentative measurements of jet rotation have been reported for jets in
different evolutionary phases from Class 0 to
T-Tauri\cite{Codella2007,Lee2008,Lee2009,Zapata2010,Chen2016,Chrysostomou2008,Bacciotti2002,Coffey2007,White2014,Coffey2015}
phase.  However, these measurements are quite uncertain because the
structures and kinematics across the jet are not well resolved.  For
example, the rotation of the T Tauri jet in DG Tau measured at $\sim$ 14 AU
spatial resolution and $\sim$ 26 \vkm{} velocity
resolution\cite{Bacciotti2002} was later called into
question\cite{White2014} through near-IR integral field observations.  Here
we report a new measurement of jet rotation in the textbook-case
protostellar jet HH 212.  Our observations with Atacama Large
Millimeter/submillimeter Array (ALMA) have a unique combination of high
spatial ($\sim$ 8 AU or \arcsa{0}{02}) and high velocity ($\sim$ 1 \vkm{})
resolution, allowing us to measure the smallest specific angular momentum in
any jet to date down to 10 AU \vkm{} scale,  as compared to
$\sim$ 30 AU \vkm{} scale in previous measurements\cite{Lee2008}.
We show that jets can indeed remove angular momentum from the innermost
disk, enabling material there to accrete toward the central protostar.









HH 212 is a nearby protostellar system deeply embedded in a compact
molecular cloud core in the L1630 cloud of Orion at a distance of $\sim$ 400
pc.  The central source is the Class 0 protostar IRAS 05413-0104, with an
estimated age of $\sim$ 40,000 yrs\cite{Lee2014}.  It has an accretion
disk\cite{Lee2017} and drives a powerful bipolar
jet\cite{Zinnecker1998}.  The jet, lying close ($\sim$ 4\degree{}) to the
plane of the sky\cite{Claussen1998}, is one of the best candidates to search
for jet rotation.  The systemic velocity is $\sim$ 1.7$\pm$1.0 \vkm{}
LSR in this system\cite{Lee2014}.




Figure~\ref{fig:jet_cont}a shows our ALMA map of the inner part of the jet
within $\sim$ 1200 AU (\arcs{3}) of the central source in SiO J=8-7 at
$\sim$ 24 AU (\arcsa{0}{06}) resolution, together with the dust continuum
map (orange image) of the accretion disk at 352 GHz\cite{Lee2017}.  SiO
J=8-7 is a dense gas tracer with a critical density of $\sim$ 10$^8$ \cmc{}
and thus traces the jet well.  This part of the jet is connected to knots
and bow structures NK, SK 1-5 seen on larger scales further
out\cite{Lee2015}.  In this map, knots SN and SS were detected before but
not resolved, and their tips were used to obtain an upper limit of jet
rotation\cite{Lee2008}.  They are now resolved.  Knot SN appears as a chain
of smaller knots and bow structures along the jet axis, as seen at
high-blueshifted velocity in Figure~\ref{fig:jet_cont}b.  Here bow
structures refer to the ones with bow wings curving backward.  These knots
and bow structures can be produced by a chain of internal shocks formed by a
semi-periodical variation in jet velocity\cite{Moraghan2016}.  Knot SS
appears as an elongated bow structure, as seen at low velocity in
Figure~\ref{fig:jet_cont}c, possibly also with a chain of smaller bow
structures along the jet axis (see also Figure~\ref{fig:jet_cont}d).  Thus,
the tips of the SN and SS knots appear to be substantially affected by shock
interaction, which makes their previous jet rotation measurements
unreliable.





Zooming closer in to the central region, a new chain of knots (N1 to N4 and
S1 to S3) are detected within $\sim$ 120 AU of the central source down to
$\sim$ 10 AU, tracing the primary jet emanating from it (see
Figure~\ref{fig:jetrot}a at $\sim$ 8 AU (\arcsa{0}{02}) resolution). 
Unfortunately, the disk is flared and optically thick\cite{Lee2017},
preventing us from detecting the jet further in.  The jet is highly
collimated with a position angle measured to be $\sim$ 23\degree{}, exactly
perpendicular to the disk major axis.  The knots are extremely narrow, with
the width (Gaussian deconvolved width, see Methods) decreasing toward the
central source, from $\sim$ 9-16 AU for knots N3 and S3 to $\lesssim$ 5-6 AU for
knots N1 and S1 (see Figure \ref{fig:jetwidthplot}).  Similar decrease in
jet width has also been seen in more evolved objects in RW Aur and DG
Tau\cite{Agra-Amboage2011}.  These widths are roughly consistent with the
width of the H$_2$O maser measured at $\sim$ 50 AU about 20 yrs
ago\cite{Claussen1998}.  For the knots further out, SS and SN, the width has
been previously estimated to be $\sim$ 80 AU at 400-800 AU in SiO
J=5-4\cite{Cabrit2007}.

Except for knot N4 that shows a bow wing on the east side of the jet axis,
the new knots do not show any hint of bow structures, allowing us to measure
radial velocity gradients with more confidence.  These knots are
closest to the central source, and thus expected to be least affected by
shocks\cite{Moraghan2016}.  Figure~\ref{fig:pvjetrot} shows the
position-velocity (PV) diagrams of the SiO emission cut across the knots
(see Figure \ref{fig:jetrot} for the cuts).  Knots N1, S1, and S2 are not
resolved, and thus no velocity gradient can be detected.  Note that for
knots S1 and N3, SiO emissions (marked with a ``?'') at ($\sim$ 5 \vkm{} LSR,
$\sim$ 4-12 AU) and ($\sim$ $-$3 \vkm{} LSR,
$\sim$ $-4$ to $-12$ AU), respectively, are offset from the rest of the SiO emission and 
are thus
likely not from the jet itself. Knots N2, N3, and S3 are marginally
resolved.  A velocity gradient can be seen across them at more than
7$\sigma$ detection (as marked by the red contours) around the mean jet
velocity (vertical dashed line).  With respect to the mean jet velocity, the
blueshifted emission is seen mostly on the west side of the knot and the
redshifted emission mostly on the east.  To measure the gradient, we
identified the emission peaks (as marked by the green squares) 
with Gaussian fits and then
fitted them with a linear velocity gradient using the least squares method. 
As marked by the solid lines, the gradients are 0.363$\pm$0.040 AU/(\vkm{})
for knot N2 over the velocity range from
$-$7.5 to 2.6 \vkm{}, 0.533$\pm$0.326 AU/(\vkm{}) for knot N3 over the
velocity range from $-$7.5 to 1.0 \vkm{}, and 0.635$\pm$0.070 AU/(\vkm{}) for
knot S3 over the velocity range from 0.12 to 8.6 \vkm{}. Since the sense of
the gradient is the same in the resolved knots, the gradients are unlikely
due to random velocity fluctuations in the jet.  The gradients are also
unlikely due to jet precession, which was found to be small with an opening
angle of $\sim$ 1\degree{} and a period of $\sim$ 100 yrs\cite{Lee2015}. 
The consistency of the sense of the gradients supports the interpretation
that they are dominated by the intrinsic jet rotation.  Moreover, this sense
of velocity gradient is the same as that for the rotation of the
disk\cite{Lee2014,Codella2014}, further supporting such an interpretation. 
Note that the rotation sense of the disk is confirmed in our new ALMA
observations of the disk in deuterated methanol at $\sim$ 16 AU resolution
down to the center\cite{Lee2017b}. The velocity gradient can be seen more
pictorially in Figure~\ref{fig:jetrot}b, where the redshifted and
blueshifted emissions are plotted separately for each knot.  For those
resolved knots, N2, N3, and S3, the redshifted emission is seen mostly on
the east and the blueshifted emission mostly on the west of the jet axis. 
The same split of emission can also be seen in the unresolved knot, N1. 
This spatial distribution of the redshifted and blueshifted emission
indicates that both the northern and southern jets are rotating clockwise
when viewed along the jet axis from north to south, in the same direction as
the disk rotation (marked by curved arrows in Figure~\ref{fig:jetrot}b). 
Assuming that the jet rotation is symmetric about the knot center, then
the resulting specific angular momentum is $l=m (\triangle V)^2/4$, where
$m$ is the gradient derived above and $\triangle V$ is the velocity range
used to derive the gradient.  Thus, the specific angular momentum is
estimated to be $\sim$ 9.4$\pm$1.0, 9.6$\pm$6.0, and 11.4$\pm$1.3 AU \vkm{}
for knots N2, N3, and S3, respectively.  Excluding the value of knot N3, which is quite
uncertain, the mean specific angular momentum is $l_j \sim 10.2\pm1.0$ AU
\vkm{}.


%





Protostellar jets are generally thought to be launched magneto-centrifugally
from disks\cite{Frank2014}.  In particular, two competing models, the X-wind
model\cite{Shu2000} and disk-wind model\cite{Konigl2000}, have been
constructed for jet launching from the accretion disks through the
magneto-centrifugal mechanism.  In this framework, the launching radius of
the jet can be derived from the specific angular momentum and the velocity
of the jet, based on conservation of energy and angular momentum, if the
mass of the central protostar is known\cite{Anderson2003} (see also
Methods).  With a protostellar mass of $M_\ast \sim$ 0.25$\pm$0.05
\solarmass{}\cite{Lee2014,Codella2014}, a jet velocity\cite{Lee2015} of $v_j
\sim$ 115$\pm$50 \vkm{}, and a mean jet specific angular momentum of
$l_j\sim$ 10.2$\pm$1.0 AU \vkm{}, the launching radius of the jet is
estimated to be $\ro\sim$ 0.05$_{-0.02}^{+0.05}$ AU, using Eq. 
\ref{eq:jradius} in Methods.

%
%



Since HH 212 has a bolometric luminosity $\sim$ 9 \solarlum{}, the dust
sublimation radius should be larger than 0.1 AU\cite{Millan2007}, which is
outside our inferred jet launching radius.  An implication is that Si is
already released from the grains into the gas phase at the base of the jet. 
However, since the jet is well collimated with a high mass-loss
rate\cite{Lee2015} of $\sim$ $10^{-6}$ \solarmass{} yr$^{-1}$, SiO is
expected to form rather quickly because of the high density in the
jet\cite{Glassgold1991}.  In addition, the Si$^+$ recombination and SiO
formation are expected to be faster than the photodissociation caused by
possible far-ultraviolet radiation of the central
protostar\cite{Cabrit2012}.

Since the SiO jet extracts the angular momentum from the innermost
region of the disk, the angular momentum problem is only partially resolved
because material still has to be transported within the disk to its
innermost region\cite{Frank2014}.  This transfer within the disk may be
achieved with other mechanisms, e.g., magneto-rotational
instability\cite{Balbus2006} and low-velocity extended tenuous disk
wind\cite{Konigl2000}.  Recent ALMA CH$_3$OH observations at $\sim$ 240
AU resolution have suggested a disk wind component in HH 212 ejected from
the disk at a radius of $\sim$ 1 AU\cite{Leurini2016}, surrounding the SiO
jet.  However, our new observations at $\sim$ 16 AU resolution show that
CH$_3$OH actually traces the disk surface within $\sim$ 40 AU of the
center\cite{Lee2017b}. Disk wind has also been suggested in other objects,
e.g., CB 26\cite{Launhardt2009}, DG Tau\cite{Agra-Amboage2011}, Orion BN/KL
Source I\cite{Greenhill1998,Matthews2010,Greenhill2013}, and
TMC1A\cite{Bjerkeli2016}.  For example, in Orion BN/KL Source I, a
low-velocity and poorly-collimated bipolar outflow was found in SiO maser to
come from the disk surface at larger disk radii of $\gtrsim$ 20 AU from the
central source\cite{Greenhill1998,Matthews2010,Greenhill2013}.  Similarly in
TMC1A, a low-velocity and poorly-collimated CO outflow was found to come
from the disk surface at larger disk radii of up to 25 AU from the
central source\cite{Bjerkeli2016}.  All these suggest the presence of a disk
wind component extracting the angular momentum from the disk at larger
radii.  In contrast, the SiO jet in HH 212 appears to come from the
innermost region of the disk, well inside 1~AU of the central star.  It is
consistent with the X-wind picture\cite{Shu2000}, but could also be the
innermost part of a more extended disk-wind\cite{Konigl2000}.

Our observations have a unique combination of high spatial and velocity
resolution, allowing us
to measure radial velocity
gradients and hence estimate the smallest specific angular momentum in
any jet to date.  The measured small specific angular momentum implies a
fast jet launched from the innermost region of the disk for the first time
in the earliest phase of star formation. Our measurement sets the tightest
constraint yet on the location of protostellar jet launching.  It also opens
up an exciting opportunity to study jet rotation and launching location in
other young systems.








\clearpage

\begin{methods}
\subsection{Observations} 


Observations of the HH 212 protostellar system were carried out with ALMA in
Band 7 at $\sim$ 350 GHz in Cycles 1 and 3, with 32-45 antennas (see Supplementary Table
\ref{tab:obs}).  The Cycle 1 project was carried out with 2 executions, both
on 2015 August 29 during the Early Science Cycle 1 phase.  The
projected baselines are 15-1466 m.  The maximum recoverable size
scale is $\sim$ \arcsa{2}{5}.  A 5-pointing mosaic was used to map the jet
within $\sim$ \arcs{15} of the central source at an angular resolution of
$\sim$ \arcsa{0}{16} (64 AU). The Cycle 3 project was carried out with 2
executions in 2015, one on November 5 and the other on December 3, during the
Early Science Cycle 3 phase.  The projected baselines are
17-16196 m.  The maximum recoverable size scale is $\sim$ \arcsa{0}{4}.  One
pointing was used to map the innermost part of the jet at an angular
resolution of \arcsa{0}{02} (8 AU). For the Cycle 1 project, the correlator
was set up to have 4 spectral windows, with one for CO $J=3-2$ at 345.795991
GHz, one for SiO $J=8-7$ at 347.330631 GHz, one for HCO$^+$ $J=4-3$ at
356.734288 GHz, and one for the continuum at 358 GHz (see Supplementary Table
\ref{tab:corr1}).  For the Cycle 3 project, the correlator was more flexible
and thus was set up to include 2 more spectral windows, with one for SO
\SOt{} at 346.528481 GHz and one for H$^{13}$CO$^+$ $J=4-3$ at 346.998338
GHz (see Supplementary Table \ref{tab:corr3}).  The total time on the HH 212 system is $\sim$ 148 minutes.

In this paper, we only present the observational results in SiO, which
traces the jet emanating from the central source.  The velocity resolution
is 0.212 \vkm{} per channel.  However, we binned 4 channels to have a
velocity resolution of 0.848 \vkm{} in order to map the jet with sufficient
sensitivities.  The data were calibrated with the CASA package
(versions 4.3.1 and 4.5) for the passband, flux, and gain (see Supplementary Table
\ref{tab:calib}).  We used a robust factor of 2 (natural weighting) for the
visibility weighting to generate the SiO maps.  In order to avoid the proper
motion effect ($\sim$ 2 AU or \arcsa{0}{005} per month using 115 \vkm{} for
the jet velocity\cite{Lee2015}), only Cycle 3 data are used to study the jet
rotation in the innermost part of the jet.  This generates a synthesized
beam with a size of \arcsa{0}{02} (8 AU) for the maps of the innermost part
of the jet (see Figure~\ref{fig:jetrot}).  In order to map the knots further
out, which are more extended, we also include the Cycle 1 data, which
has a larger maximum recoverable scale. In addition, a taper of
\arcsa{0}{05} was used to degrade the beam size to \arcsa{0}{06} (24 AU, see
Figure~\ref{fig:jet_cont}) in order to improve the S/N ratio. The
noise levels can be measured from line-free channels and are found to
be $\sim$ 1.6 \mJyb{} (or $\sim$ 40 K) for a beam of $\sim$ \arcsa{0}{02}
(8 AU) and 1.9 \mJyb{} (or $\sim$ 6 K) for a beam of $\sim$
\arcsa{0}{06} (24 AU), respectively. The velocities in the channel
maps and the resulting position-velocity diagrams are LSR.




\subsection{Gaussian deconvolved width of the jet knots:}

Supplementary Figure \ref{fig:jetwidth} shows the spatial profile of the jet knots
perpendicular to the jet axis, extracted from the SiO total intensity map
shown in Figure \ref{fig:jetrot}a (see the white lines for the cuts).  In
order to derive the width of the knots, we fitted the spatial profiles of
the knots with a gaussian profile.  For knot S1, the emission at $\sim$ 4-12
AU is unlikely from the jet itself, as discussed in the main text, and is
thus excluded from the fitting.  The deconvolved width is the width
deconvolved with the beam size of $\sim$ 8 AU (\arcsa{0}{02}).  Knots N2,
N3, and S3 have a deconvolved width greater than the beam size.
Knots N1, S1, and S2 have a deconvolved width smaller
than the beam size.


\def\no{n_\mathrm{o}}
\def\na{n_\mathrm{t}}
\def\Ro{R_\mathrm{o}}
\def\Ra{R_\mathrm{t}}
\def\To{T_\mathrm{o}}
\def\Ta{T_\mathrm{t}}
\def\ho{h_\mathrm{o}}
\def\ha{h_\mathrm{t}}

\def\vk{v_\mathrm{ko}}
\def\cs{c_\mathrm{s}}
\def\vko{v_\mathrm{ko}}
\def\cso{c_\mathrm{so}}
\def\vp{v_\phi}

\subsection{Mean (or Systemic) Velocities of the Jet:} 

%

Supplementary Figure~\ref{fig:pvjet} shows the position-velocity diagram of the SiO jet
cut along the jet axis.  The northern jet component is detected from $\sim$
$-$14 to 8 \vkm{} LSR, with a mean velocity of $\sim$$-$3 \vkm{} LSR (as
indicated by the vertical dashed line).  The southern jet component is
detected from $\sim$ $-$5 to 13 \vkm{}  LSR, with a mean velocity of
$\sim$ 4 \vkm{} LSR (as indicated by the vertical dashed line).  These
mean velocities are taken to be the systemic velocities in the northern and
southern jet components.

\subsection{Estimation of Jet Launching Radius}

Protostellar jets are generally thought to be launched magneto-centrifugally
from disks\cite{Frank2014}.  In this framework, the launching radius of the
jet can be derived from the specific angular momentum and the velocity of
the jet, based on conservation of energy and angular momentum along the
field line, if the mass of the central protostar is
known\cite{Anderson2003}.  For HH 212, since (1) the jet velocity (poloidal
velocity) is so high that the gravitational potential can be neglected at
large distances, (2) the jet velocity is much higher than the jet rotation,
and (3) the jet inclination angle is very small ($\sim$
4\degree{})\cite{Claussen1998}, the governing equation (Eq.  4 in Anderson
et al.  2003)\cite{Anderson2003} that is used to derive the jet launching
radius can be simplified and rewritten as
\begin{equation}
\frac{2 l_j}{v_j^2}\Big(\frac{GM_\ast}{\ro}\Big)^{1/2} - \frac{3GM_\ast}{v_j^2}-\ro \approx 0
\end{equation}
in order find approximate solutions analytically, where $\ro$ is the
launching radius at the footpoint, $v_j$ is the jet velocity, $l_j$ is the
specific angular momentum of the jet measured at a large distance, and
$M_\ast$ is the mass of the central protostar.  Solving this equation, we
find the jet launching radius to be \begin{equation}
\ro \approx \Big(\frac{2 l_j}{v_j^2}\Big)^{2/3}(GM_\ast)^{1/3}\Big[1-\frac{2}{3}\eta + \frac{1}{9} \eta^2\Big]
\label{eq:jradius}
\end{equation}
with 
\begin{equation}
\eta = \frac{3}{2^{2/3}}(\frac{GM_\ast}{v_j l_j})^{2/3}
\end{equation}
Note that the second and third terms are the correction terms to the
previous solution (Eq.  5 in Anderson et al. 2003)\cite{Anderson2003}.
They can improve the accuracy of the launching radius estimate in the case
where the dimensionless parameter $\eta$ is not much smaller than unity,
particularly when the specific angular momentum $l$ is relatively small, as
is true for HH212.

\end{methods}

\clearpage
\newcommand\aap{{A\&A}}
\newcommand\apjl{{ApJL}}
\newcommand\apj{{ApJ}}
\newcommand\apjs{{ApJS}}
\newcommand\aj{{AJ}}
\newcommand\araa{{ARAA}}
\newcommand\nat{{Nature}}
\newcommand\mnras{{MNRAS}}

\begin{addendum}

\item[Correspondence and Requests for Materials]  should be addressed to
Chin-Fei Lee.

\item This paper makes use of the following ALMA data:
ADS/JAO.ALMA\#2012.1.00122.S and 2015.1.00024.S.  ALMA is a partnership of
ESO (representing its member states), NSF (USA) and NINS (Japan), together
with NRC (Canada), NSC and ASIAA (Taiwan), and KASI (Republic of Korea), in
cooperation with the Republic of Chile.  The Joint ALMA Observatory is
operated by ESO, AUI/NRAO and NAOJ.  C.-F.L.  acknowledges grants from the
Ministry of Science and Technology of Taiwan (MoST 104-2119-M-001-015-MY3)
and the Academia Sinica (Career Development Award).  ZYL is supported in
part by NASA NNX14AB38G and NSF AST 1313083.

\item[Author Contributions]

C.-F. Lee led the project, analysis, discussion, and drafted the manuscript. 
P.T.P.  Ho and Z.-Y.  Li commented on the manuscript and participated in the
discussion.  All other coauthors contribute to scientific discussion.


\item[Data Availability Statement]

This letter makes use of the following ALMA data:
ADS/JAO.ALMA\#2012.1.00122.S and 2015.1.00024.S.  The data that support the
plots within this paper and other findings of this study are available from
the corresponding author upon reasonable request.

\item[Competing interests]
The authors declare no competing financial interests.

\end{addendum}
  
\clearpage

 
\begin{figure}
\centering
\putfig{0.7}{270}{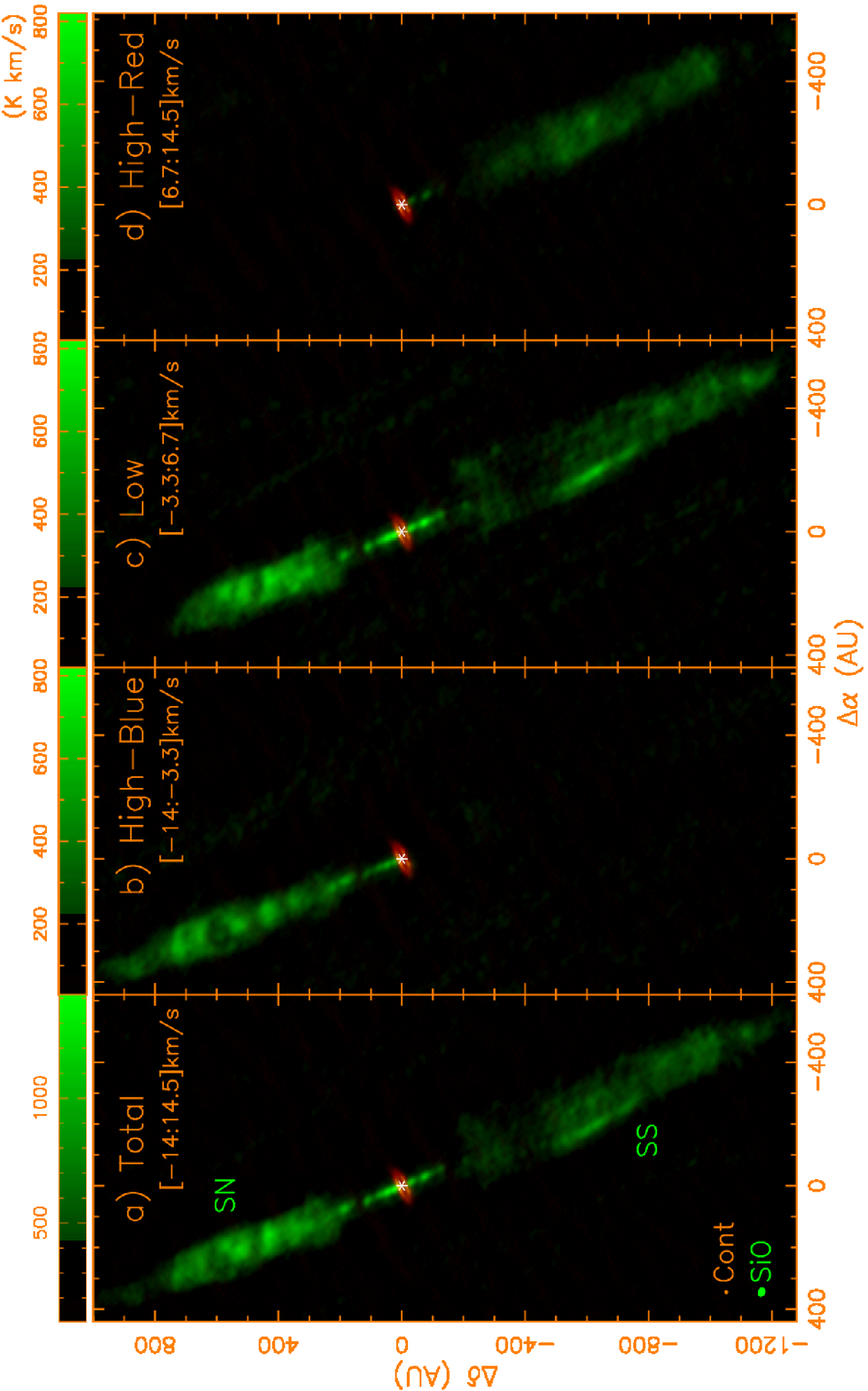} 
\end{figure}
\begin{mfigure}
{ALMA SiO J=8-7 maps (green images) of the jet at a spatial resolution
of $\sim$ 24 AU (\arcsa{0}{06}), on top of the dust continuum map (orange image) of
the accretion disk at 352 GHz\cite{Lee2017}.  The velocity ranges used for
the SiO emission are indicated in the brackets in the upper right corner.
The systemic velocity is assumed to be 1.7$\pm$0.1 \vkm{} LSR, as found before\cite{Lee2014}.
Velocities within 5 \vkm{} of the systemic velocity are
referred to as low and those outside the range as high.
Asterisks mark the source position at
$\alpha_{(2000)}=\ra{05}{43}{51}{4086}$,
$\delta_{(2000)}=\dec{-01}{02}{53}{147}$\cite{Lee2017}.
\label{fig:jet_cont}}
\end{mfigure}

\begin{figure}
\centering
\putfig{0.8}{270}{f2.eps} 
\end{figure}
\begin{mfigure}
{A zoom-in to the innermost part of the jet in SiO within $\sim$ 120 AU (\arcsa{0}{3}) of the central source,
at an angular resolution of $\sim$ 8 AU (\arcsa{0}{02}),
on top of the continuum map of the disk. 
The maps show the intensity (in unit of K \vkm{}) integrated over
certain velocity range as given in the later part of the caption.
White lines mark the cuts used to extract
the spatial profiles and the position-velocity diagrams of the knots, with
dots being the centers. 
(a) A chain of new knots (N1..N4, S1..S3) are detected, tracing the primary jet emanating
from the disk. The SiO is integrated from $\sim$ $-$14 to 14.5 \vkm{} LSR.
Contour levels start from 3$\sigma$ with a step of 1.5$\sigma$, where $\sigma=330$ K \vkm{}.
(b) Blueshifted and redshifted
SiO emission of the jet plotted with the continuum emission.
In the north, the velocity ranges of the blueshifted and redshifted emission are
$\sim$ $-$10 to $-$5 \vkm{} LSR and $-$1 to 4 \vkm{} LSR, respectively.
In the south, the velocity ranges of the blueshifted and redshifted emission are
$\sim$ $-$3 to 2 \vkm{} LSR and 6 to 11 \vkm{} LSR, respectively.
The direction of disk rotation is depicted by curved arrows.
\label{fig:jetrot}}
\end{mfigure}

\begin{figure}
\centering
\putfig{1}{270}{f3.eps} 
\end{figure}
\begin{mfigure}
{The jet (gaussian deconvolved) width measured for the new knots within
$\sim$ 100 AU of the source. The ``star" marks the H$_2$O maser width
measured by VLBI about 20 yrs ago\cite{Claussen1998}. The error bars
show the uncertainties in the width. The down arrows indicate that the measured
widths are the upper limits.
\label{fig:jetwidthplot}}
\end{mfigure}

\begin{figure}
\centering
\putfig{0.65}{270}{f4.eps} 
\end{figure}
\begin{mfigure}
{Position-velocity (PV) diagrams cut across the knots (N1..N3 and S1..S3) in the jet.
The horizontal dashed lines indicate the peak (central) position of the knots.
The vertical dashed lines indicate roughly the systemic (mean) velocities
for the northern and southern jet components (as in Figure~\ref{fig:pvjet}, see Methods).
The contour levels start from 4$\sigma$ with a step of 1$\sigma$, where $\sigma=21.3$ K.
The red contours mark the 7$\sigma$ detections in knots N2, N3, and S3.
For knots N3 and S1, the emissions marked with ``?'' are likely not from the jet itself.
The green squares mark the emission peak positions with more than 7$\sigma$ detections,
as determined from the Gaussian fits. The error bars show the uncertainties in the
peak positions, which are assumed to be given by a quarter of the FWHM.
The solid lines mark the
linear velocity structures across the knots. The bars indicate the angular (8 AU or
$\sim$ \arcsa{0}{02}) and velocity ($\sim$ 1.7 \vkm{}) resolutions used
for the PV cuts.
\label{fig:pvjetrot}}
\end{mfigure}

\end{document}